\documentclass[aps,prl,reprint,superscriptaddress]{revtex4-1}

\usepackage{graphicx}
\usepackage[utf8]{inputenc}
\makeatletter
\def\blfootnote{\xdef\@thefnmark{}
\@footnotetext}
\makeatother
\begin{document}


\title{Impact of $^{16}$O($e,e'\alpha$)$^{12}$C measurements on the $^{12}$C($\alpha,\gamma$)$^{16}$O astrophysical reaction rate} 


\author{R. J. Holt}
\affiliation{Physics Division, Argonne National Laboratory, Argonne, Illinois 60439}
\affiliation{Kellogg Radiation Laboratory, California Institute of Technology, Pasadena, California 91125}
\email{rholt@caltech.edu}
\author{B. W. Filippone}
\affiliation{Kellogg Radiation Laboratory, California Institute of Technology, Pasadena, California 91125}
\email{bradf@caltech.edu}



\date{\today}

\begin{abstract}
The $^{12}$C($\alpha,\gamma$)$^{16}$O reaction, an important component of stellar helium burning, has a key role in nuclear astrophysics. It has direct impact on the evolution and final state of massive stars and also influences the elemental abundances resulting from nucleosynthesis in such stars.  Providing a reliable estimate for the energy dependence of this reaction at stellar helium burning temperatures has been a longstanding and important goal. In this work, we study the role of potential new measurements of the reaction, $^{16}$O($e,e'\alpha$)$^{12}$C reaction, in reducing the overall uncertainty. A multilevel $R$-matrix analysis is used to make extrapolations of the astrophysical S factor for the $^{12}$C($\alpha,\gamma$)$^{16}$O reaction to the stellar energy of 300 keV. The statistical precision of the $S$-factor extrapolation is determined by performing multiple fits to existing $E1$ and $E2$ ground state capture data, including the impact of possible future measurements of the $^{16}$O($e,e'\alpha$)$^{12}$C reaction.  In particular, we consider a proposed MIT experiment that would make use of a high-intensity low-energy electron beam that impinges on a windowless oxygen gas target as a means to determine the total $E1$ and $E2$ cross sections for this reaction. 
\end{abstract}


\maketitle

\section{Introduction}
The $^{12}$C($\alpha,\gamma$)$^{16}$O reaction is one of the most significant reactions in nuclear astrophysics\cite{Fowler:1984zz,Woosley:2003nki}. A recent review\cite{deBoer:2017ldl} illustrates the importance of this reaction in both the evolution of and nucleo-synthetic yields from massive stars. The purpose of this study is to explore the role that forthcoming measurements of the  $^{16}$O($e,e'\alpha$)$^{12}$C  (OSEEA)  reaction could have on reducing the overall uncertainty in the cross section for the $^{12}$C($\alpha,\gamma$)$^{16}$O reaction at helium burning temperatures. To do this, we follow the procedure given in Ref.~\cite{Holt:2018tbi,Holt:2018cgp}.  We perform fits to the existing data using the $R$-matrix approach\cite{Lane:1948zh} and study the impact of including the planned new data on the statistical error. This is achieved by starting with a reasonable $R$-matrix fit that can be used as a basis for comparison to fits with and without projected OSEEA data.  In particular, we consider a proposed MIT experiment\cite{Friscic:2019eow} in order to assess the possible role of new measurements in reducing the overall uncertainty in the cross section.  In this work the $E1$ and $E2$ $^{12}$C($\alpha,\gamma$)$^{16}$O ground state cross sections can be extracted\cite{Friscic:2019eow} from measurements of the OSEEA.  A detailed $R$-matrix analysis of the $^{12}$C($\alpha,\gamma$)$^{16}$O (CTAG) reaction and an excellent review of the subject is given in ref.\cite{deBoer:2017ldl}.  

In the present work, we employed the $R$-matrix approach to calculate the total cross section, $\sigma(E)$, for CTAG to the ground state.  Considering only ground state capture is sufficient for this study since the capture to excited states is believed\cite{deBoer:2017ldl} to contribute only about 5$\%$ to the total capture rate at 300 keV.  The cross section is then used to calculate the astrophysical $S$ factor given by

\begin{equation}
S(E) = \sigma(E)Ee^{2\pi \eta}
\end{equation}
where $E$ is the energy in the center of mass, $\eta$ is the Sommerfeld parameter, $\sqrt{\frac{\mu}{2E}}Z_1Z_2\frac{e^2}{\hbar}$, and $\mu$ is the reduced mass of the carbon ion and alpha particle.   For the $^{12}$C($\alpha,\gamma$)$^{16}$O reaction, the value of $S$ at $E=300\ keV$ is typically quoted as this is the most probable energy for stellar helium burning.     We performed extrapolations to 300 keV in order to study the impact of potential new data.  Efforts aimed at improving the data and extrapolation are underway\cite{suleiman:2014aa,Gai:2018sip,Balabanski:2017qth,Costantini:2009wn,Robertson:2016llv,Liu:2017zvh,Bemmerer:2018zsh,xu:2007,Friscic:2019eow} at a number of laboratories worldwide.   The new inverse reaction $^{16}$O($\gamma,\alpha$)$^{12}$C (OSGA) experiments\cite{suleiman:2014aa,xu:2007,Gai:2018sip,Balabanski:2017qth} as well as the forthcoming OSEEA reaction\cite{Friscic:2019eow} bring different sets of systematic errors than previous experiments and thus provide an additional check on systematics.

\section{$R$-matrix fits and $S$-factor projections}

The R-matrix analysis and equations used here have already been described in Ref.~\cite{Holt:2018tbi}.
As before, we only considered ground state transitions and statistical errors in this study.  We chose a channel radius of 5.43 fm to be consistent with a previous analysis\cite{deBoer:2017ldl}
 We employed five E1 resonance levels and four E2 resonance levels in the internal part of the the R-matrix analysis as shown in Table~ \ref{one}.  The parameters in Table~\ref{one} are defined in Ref.~\cite{Holt:2018tbi}.   Because both $E1$ and $E2$ $S$ factors can be determined from OSEEA, unlike the proposed JLab experiment, the values of the fitted parameters in Table~\ref{one} are slightly different from those of Ref.~\cite{Holt:2018tbi}.  As before we turned off the external part for this study in order to speed up computations.  


\begin{table}[!htp] 
\caption{\label{one}Best fit parameters used in the present fits to the CTAG data for $E1$ and $E2$ separately, and a channel radius of 5.43 fm. 
 The widths for resonances above threshold are the observable widths $\Gamma_{\lambda\alpha}$.  The widths for the bound states are reduced widths $\gamma^2_{1\alpha b}$.  The minus signs in front of the widths indicate the signs of the reduced width amplitudes.  The values marked with an asterisk were allowed to vary in the fit, and are given for the ``all" fit in Table~\ref{two}.  All other parameters were fixed.  The parameters are defined in Ref.~\cite{Holt:2018tbi}}.
\begin{ruledtabular}
\begin{tabular}{c c c c  c c c}
   &    &    $E1$   &    &         &   $E2$  & \\
$\lambda$ &  $E_\lambda$  & $\Gamma_{\lambda\alpha }/\gamma^2_{1 \alpha  b}$  & $\Gamma_{\lambda\gamma \circ}$  &E$_\lambda$ &  $\Gamma_{\lambda\alpha}/\gamma^2_{1\alpha  b}$  & $\Gamma_{\lambda\gamma \circ}$  \\
   &   (MeV) & (keV) & (eV) & (MeV) & (keV) & (eV)\\
\hline
1  &   -0.305 &  118.3$^*$            &  0.055          &   -0.480    &   104.1$^*$    &  0.097  \\
2  &    2.416  & 396.9$^*$            &  -0.0146$^*$    &    2.683    &   0.62            &   -0.0057\\
3  &    5.298  &  99.2                  &  5.6              &    4.407      &   83.0          &  -0.65  \\
4  &    5.835  &  -29.9                 &  42.0                &    6.092        &   -349          &  -0.911$^*$  \\
5  &   10.07     & 500                    &  0.522$^*$        &     -           &    -               &   -   \\
\end{tabular}
\end{ruledtabular}
 \end{table}


\begin{figure}[!hb]
\includegraphics[width=3.4in]{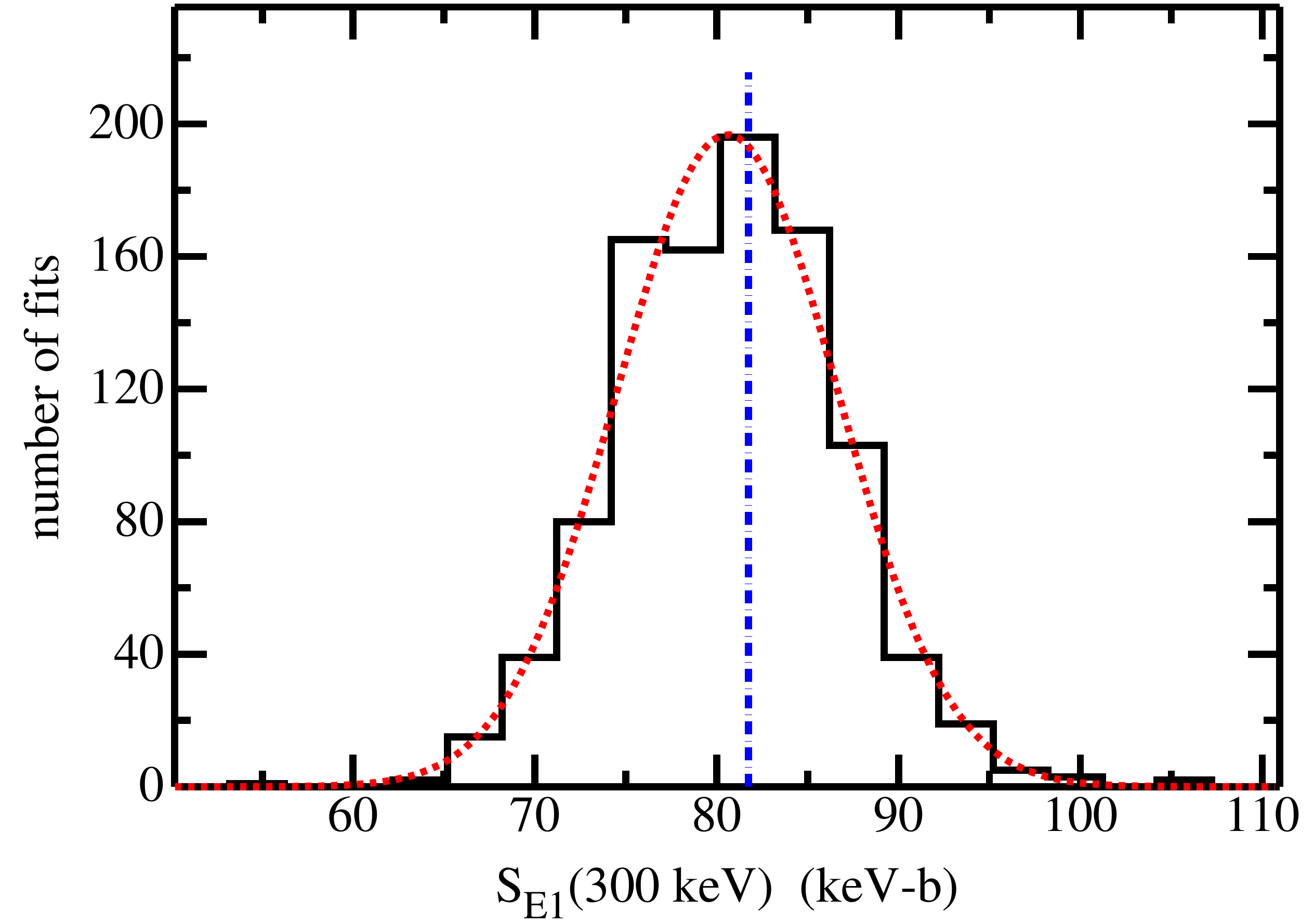}
\includegraphics[width=3.4in]{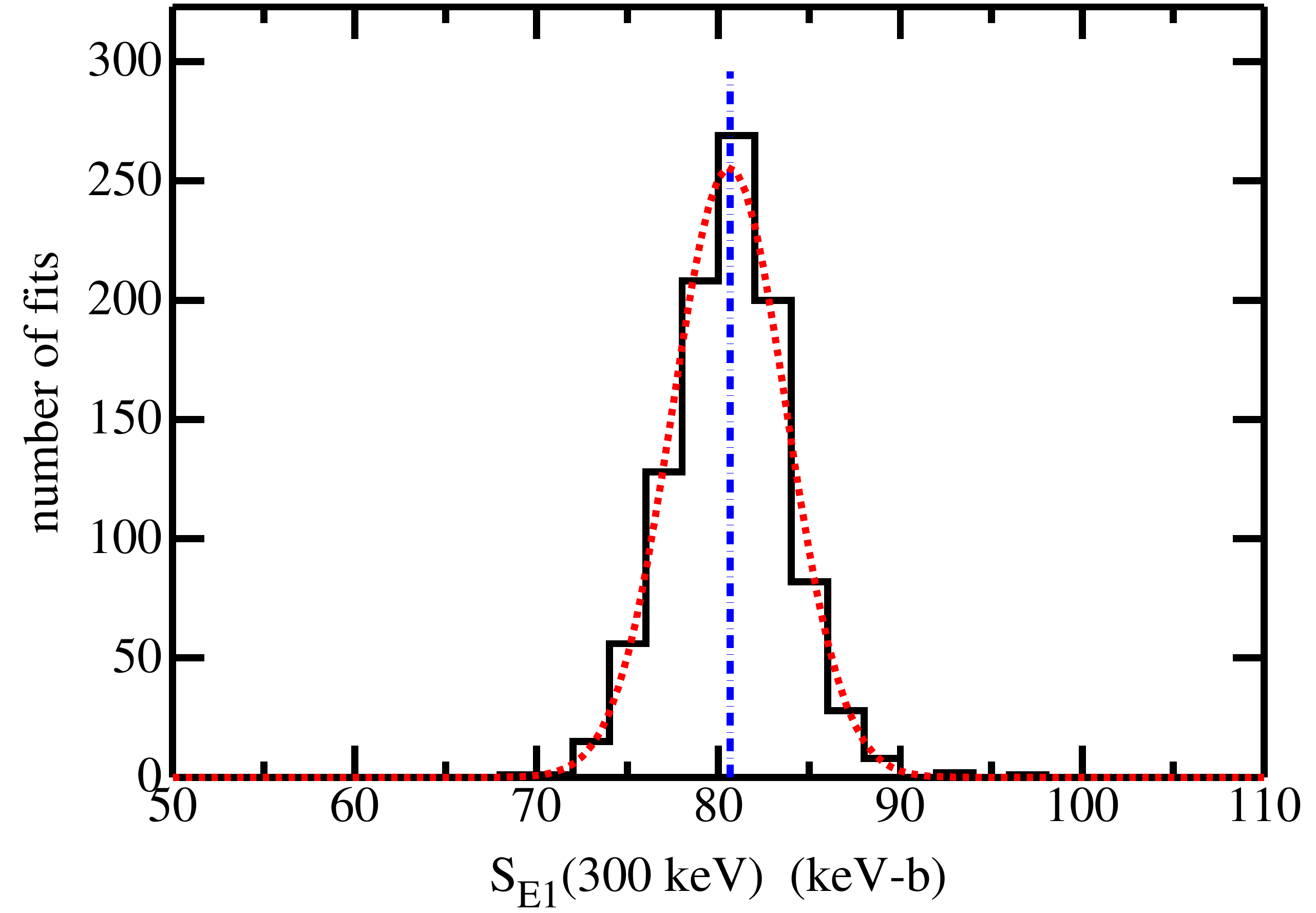}
\caption{Projections of the astrophysical $S_{E1}$ factor to 300 keV for fits of existing $E1$ data (top panel) and for existing $E1$ plus proposed OSEEA data (bottom) for a channel radius of 5.43 fm. The blue dashed vertical lines indicate the projections for the fit to the original data, while the histograms represent the results of 1000 fits to randomized pseudo-data that would lie along the fit to original data.  The red dotted curves are Gaussians based on the means and standard deviations found from the fits. }
\label{fig1}
\end{figure}
\begin{figure}[]
\includegraphics[width=3.4in]{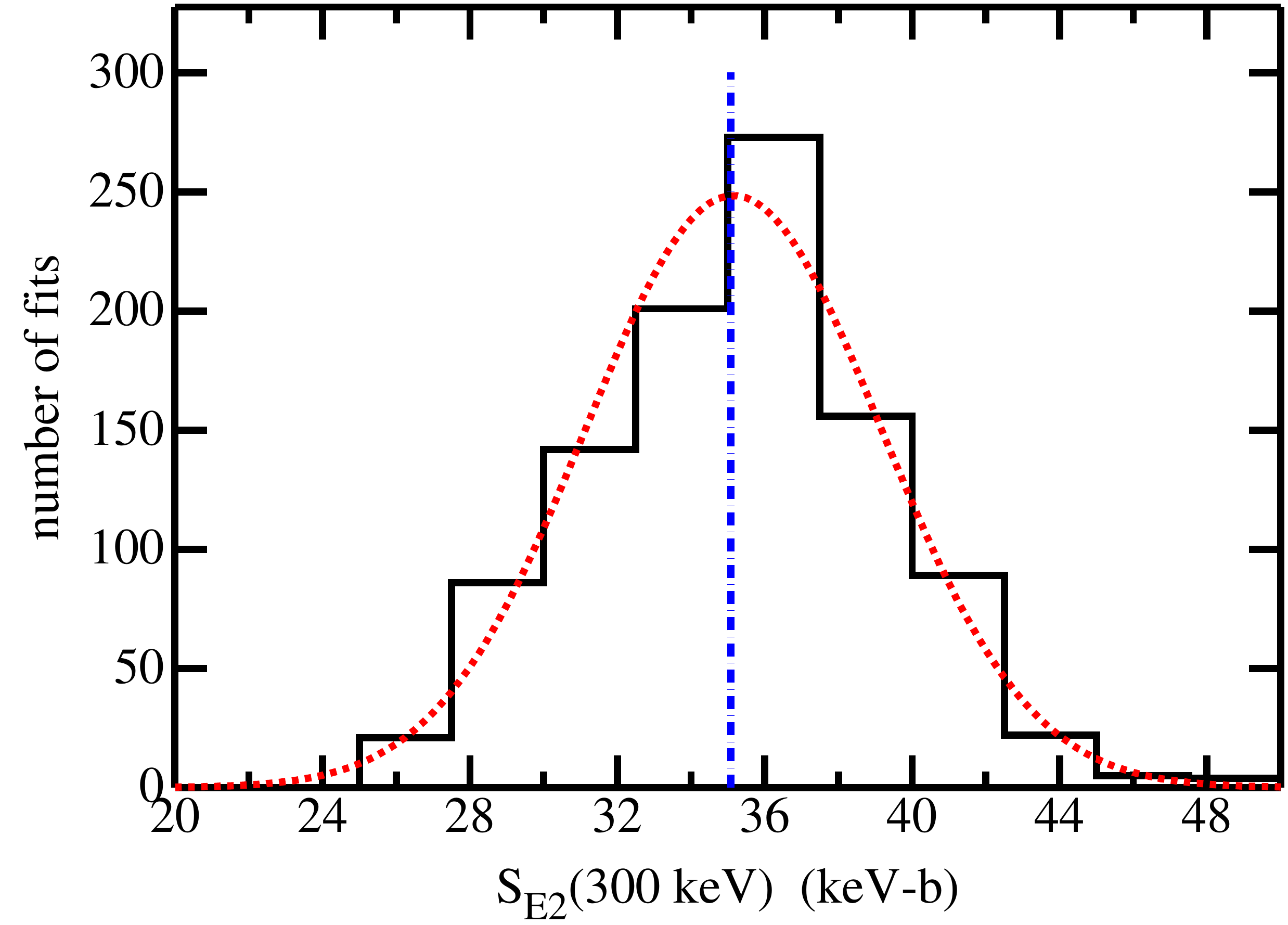}
\includegraphics[width=3.4in]{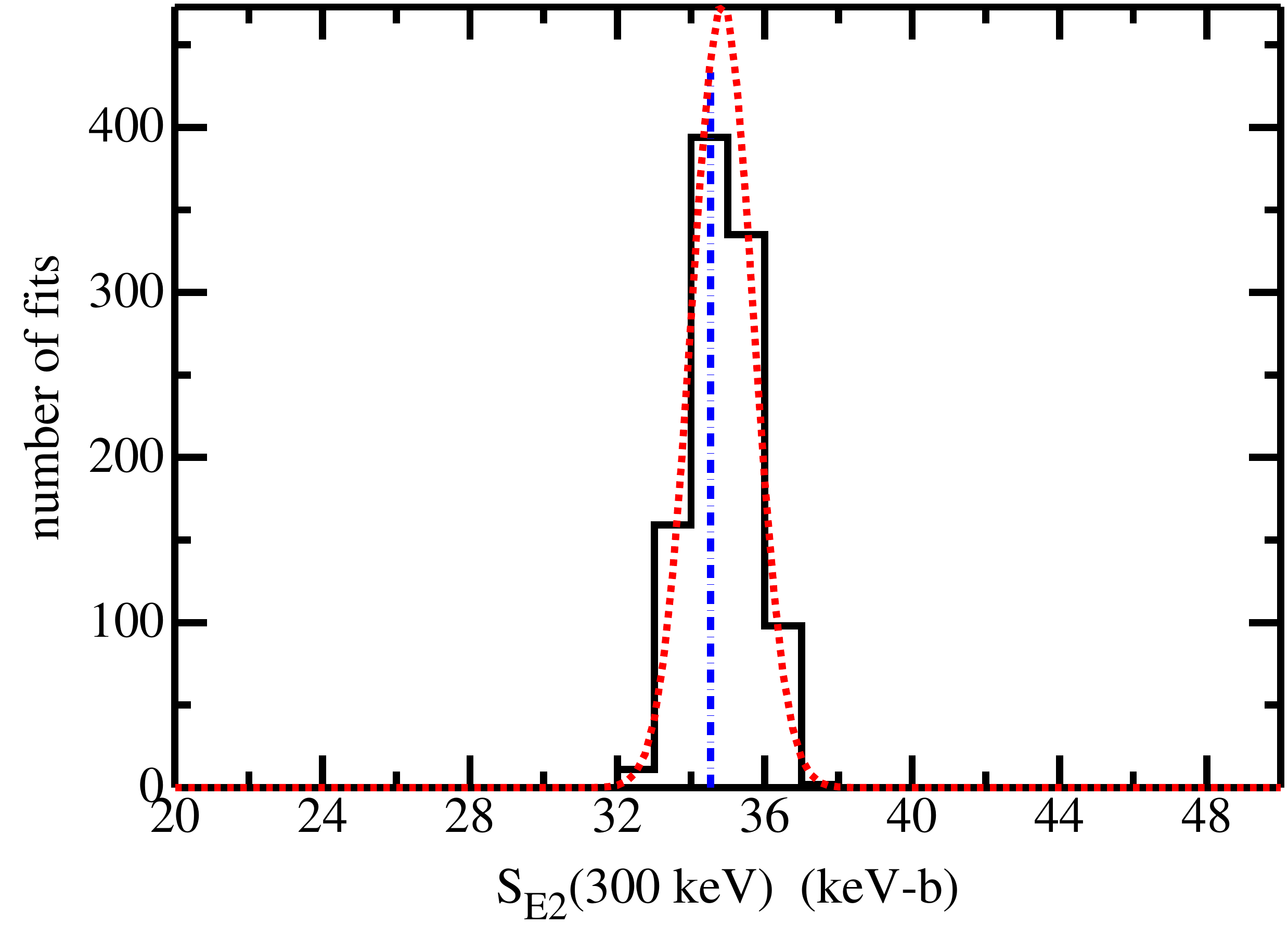}

\caption{Projections of the astrophysical $S_{E2}$ factor to 300 keV for fits of existing $E2$ data (top panel) and for existing $E2$ plus proposed MIT data (bottom) for a channel radius of 5.43 fm. The blue dashed vertical lines indicate the projections for the fit to the original data, while the histograms represent the results of 1000 fits to randomized pseudo-data that would lie along the fit to original data.  The red dotted curves are Gaussians based on the means and standard deviations found from the fits. }
\label{fig2}
\end{figure}

We used a SIMPLEX fitter\cite{Nelder:1965zz} for the present work.  Our best R-matrix fits of the existing $E1$ and $E2$ $S$ factor data were taken as the most probable descriptions of the $S_{E1}$ and $S_{E2}$ factor data.  In order to explore the statistical variation in the $S$-factor extrapolations, we created $S_{E1}$ and $S_{E2}$  pseudo-data for the existing CTAG data by random variation according to a Gaussian probability distribution about the best fit $S_{E1}$ and $S_{E2}$ values at the measured energies.  In the randomizations, we multiplied the individual pseudo-data uncertainties as taken from Ref.~ \cite{Friscic:2019eow,friscic} by the square root of the ratio of the original best fit $S$ factor values, defined by the fit values in Table~\ref{one} to the original measured uncertainties, given by the CTAG experiments.  We further multiplied these uncertainties by the square root of the $E1$ and $E2$ reduced chi squares, the Birge factor\cite{birge}, for the $E1$ and $E2$ fits, respectively.  This procedure should give a conservative estimate for statistical uncertainties of the extracted values for $S_{E1}$ and $S_{E2}$.  For the subtheshold states, we fixed the radiative widths of the subthreshold states at the measured values and varied the reduced alpha widths.  We allowed the reduced alpha and radiative width of the first $E1$ state above threshold to vary in the fit, while we allowed the radiative width of the fifth $E1$ state to vary.   We also allowed the radiative width of the fourth $E2$ $R$-matrix level to vary.
 The first $E2$ state above threshold is very narrow and we fixed the parameters of this level at those of ref.{\cite{deBoer:2017ldl}.  The radiative width of the third $E2$ resonance was treated separately.  We observed that using the value in ref.{\cite{deBoer:2017ldl} resulted in a cross section that was significantly smaller than the data of ref.\cite{Schurmann:2011zz}.  Rather, we made a fit to $E2$ data that included the data of ref.\cite{Schurmann:2011zz}.  We then fixed the third $E2$ radiative width at -0.65 eV found from the fit and used it in subsequent fits to the data below 3 MeV.  Indeed, we fixed all other parameters except the third $E2$ radiative width and those marked with an asterisk in table~\ref{one} at the values of ref.{\cite{deBoer:2017ldl}.

 Also, following ref. {\cite{deBoer:2017ldl}}, we performed the fits by maximizing L rather than minimizing $\chi^2$, where L is given\cite{sivia} by
\begin{equation}
L = \Sigma_i ln[(1-exp(-R_i/2))/R_i]
\end{equation}
and $R_i=(f(x_i)-d_i)^2/\Delta S_i^2$ is the usual quantity used in $\chi^2$ minimizations.  Here $f(x_i)$ is the function to be fitted to data, $d_i$, with statistical error $\Delta S_i$.  The L maximization has the feature that it reduces the impact of large error bar data on the fit and generally gives larger S-factor uncertainties in projected values of $S(300\ keV)$ than that of a $\chi^2$ minimization.   This work differs slightly from that in Ref.~\cite{Holt:2018tbi} in that we maximized $L_{E1}$ and $L_{E2}$ separately in this work
where $L_{E1(2)}$ is $L$ for $E1(2)$ data both with and without the possible forthcoming OSGA reaction data or OSEEA data in this case.  This leads to slightly different fit parameters in Table~\ref{one} compared with those in Ref.~\cite{Holt:2018tbi}.  The parameters in Table~\ref{one} result in an $S_{E1}$(300 keV) and $S_{E2}$(300 keV) given by the blue dash-dot line in Fig.~\ref{fig1} and Fig.~\ref{fig2}.

The parameters of the bound levels are very important for the projection to 300 keV. The resonance energies were fixed, but the parameters, $E_\lambda$, depend on the reduced width of the levels. We allowed the reduced widths of the bound states to vary, so the $E_\lambda$ varies.  We chose the R-matrix boundary condition constants to cancel out this effect for the second levels so that the $E_\lambda$ are the resonance energies for these levels. For the third and higher levels, the reduced widths were not varied because alpha elastic scattering determined these widths and allowing them to vary did not make a significant difference. 
 We used the CTAG $S$-factor data sets given in refs. \cite{Dyer:1974pgc,Kremer:1988zz,Redder:1987xba,Ouellet:1992zz,Roters:1999zz,Gialanella:2001ayx,Kunz:2001zz,Assuncao:2006vy,Makii:2009zz,Plag:2012zz}.



Proposed OSGA experiments\cite{suleiman:2014aa,Ugalde:2012eh,DiGiovine:2015lda,Gai:2018sip,Balabanski:2017qth} as well as the OSEEA experiment\cite{Friscic:2019eow} are expected to have several orders of magnitude improvement in integrated luminosity over previous experiments and should provide data at the lowest practical values of energy.  We take our best R-matrix fits to the $E1$ and $E2$ CTAG $S$-factor data as the most probable description of the projected MIT data\cite{Friscic:2019eow}.  We then randomly varied the OSEEA $S_{E1}$ and $S_{E2}$-factor pseudo-data based on their projected uncertainties\cite{Friscic:2019eow} according to a Gaussian probability distribution about the best fit $S_{E1}$ and $S_{E2}$-factor values. 
In order to study the impact of proposed OSEEA data and low energy data in particular, we performed four fits:  a fit to all existing $E1$ and $E2$ data separately (denoted by ``all" in Table~\ref{two}); a fit to data published after the year 2000 (denoted by ``2000"), both with (denoted by ``M" in Table~\ref{two}) and without projected MIT data. 
Although it has been customary\cite{Perez:2016aol} to eliminate data sets that deviate by more than three standard deviations from the fitted results, we chose to select data sets after the year 2000 as a test of systematic deviations and as suggested by Strieder\cite{strieder}.  This approach assumes that experimental equipment and methods have improved over the decades.  Another reason for this approach is that not all authors of the data sets disclose their systematic errors.  The $S$ factors projected to 300 keV along with standard deviations, $\Delta S$, which represent the statistical fit uncertainty are given in Table~\ref{two} for the four cases.  The reduced $\chi^2$ for the fit to the initial data is also shown.  

\begin{table}[!ht] 
\caption{\label{two}S-factor projections to 300 keV and standard deviations for total $S$, $S_{E1}$ and $S_{E2}$ for fits with a channel radius of 5.43 fm.  The data choices are defined in the text.}
\begin{ruledtabular}
\begin{tabular}{l c c c c  c c c c}
data & init $\chi_\nu^2 $& init $\chi_\nu^2$ &$ S$  & $\Delta S$  & $S_{E1}$  &   $\Delta S_{E1}$  &  $S_{E2}$  &   $\Delta S_{E2}$  \\
        &    $E1$ & $E2$ 	  &  &  &  &  (keV-b)    \\
\hline
all            &    2.6 & 1.5      &   116.8         &     7.3     &   81.7        &     6.1      &    35.1     &   4.0  \\
all M         &    2.7 & 1.6      &   115.2          &     3.2     &   80.7       &    3.1       &     34.5    &    0.8  \\
2000        &     1.7     &  1.9  &  115.3          &     8.3     &    79.3      &    7.0       &    36.0     &   4.4   \\
2000 M     &     1.4  &  2.0     &  117.4          &     4.3     &    82.9      &    4.2       &     34.5    &   0.9   \\
\hline
\end{tabular}
\end{ruledtabular}
 \end{table}



\begin{figure}[]
\begin{center}
\includegraphics[width=3.4in]{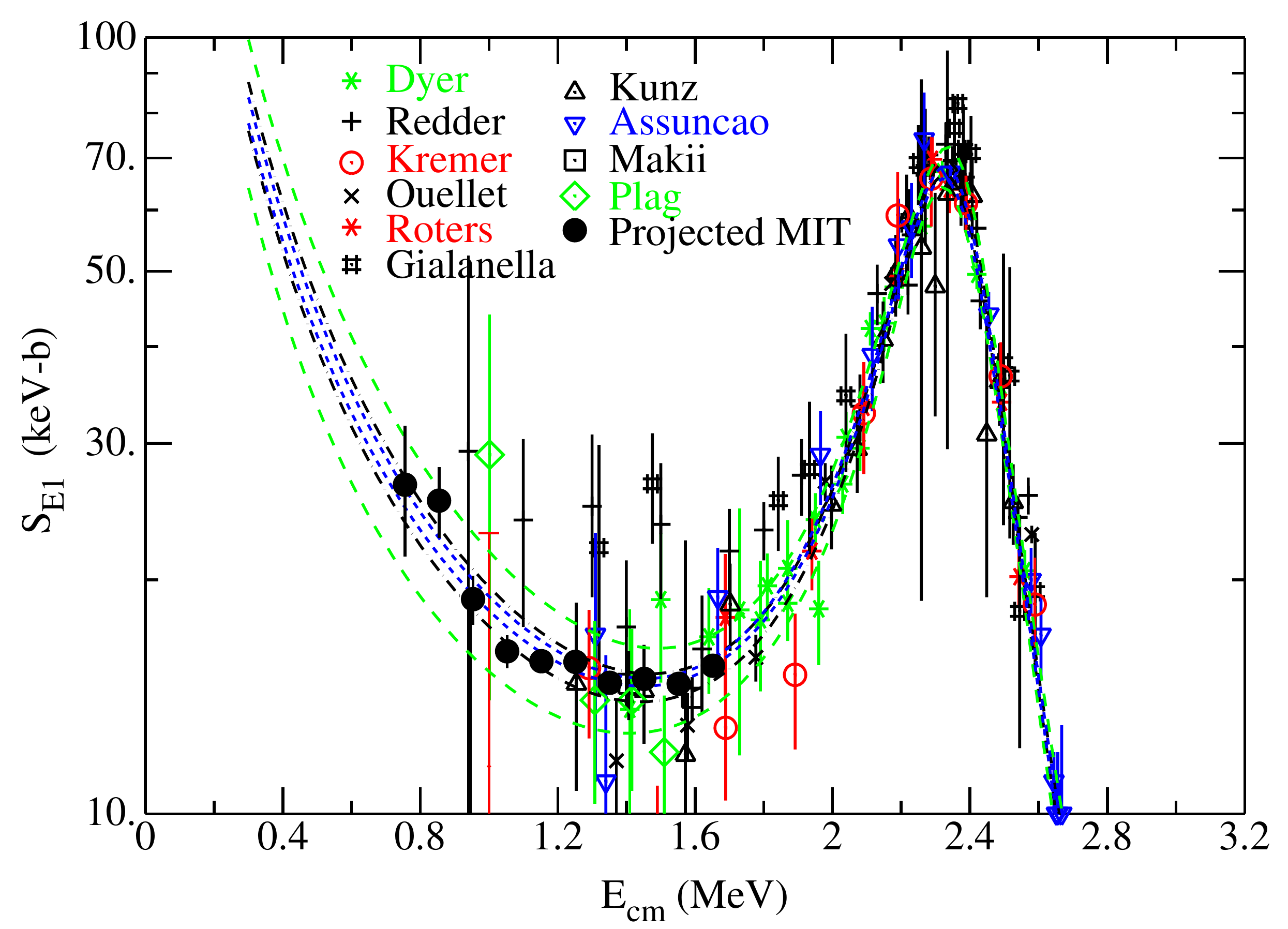}
\includegraphics[width=3.4in]{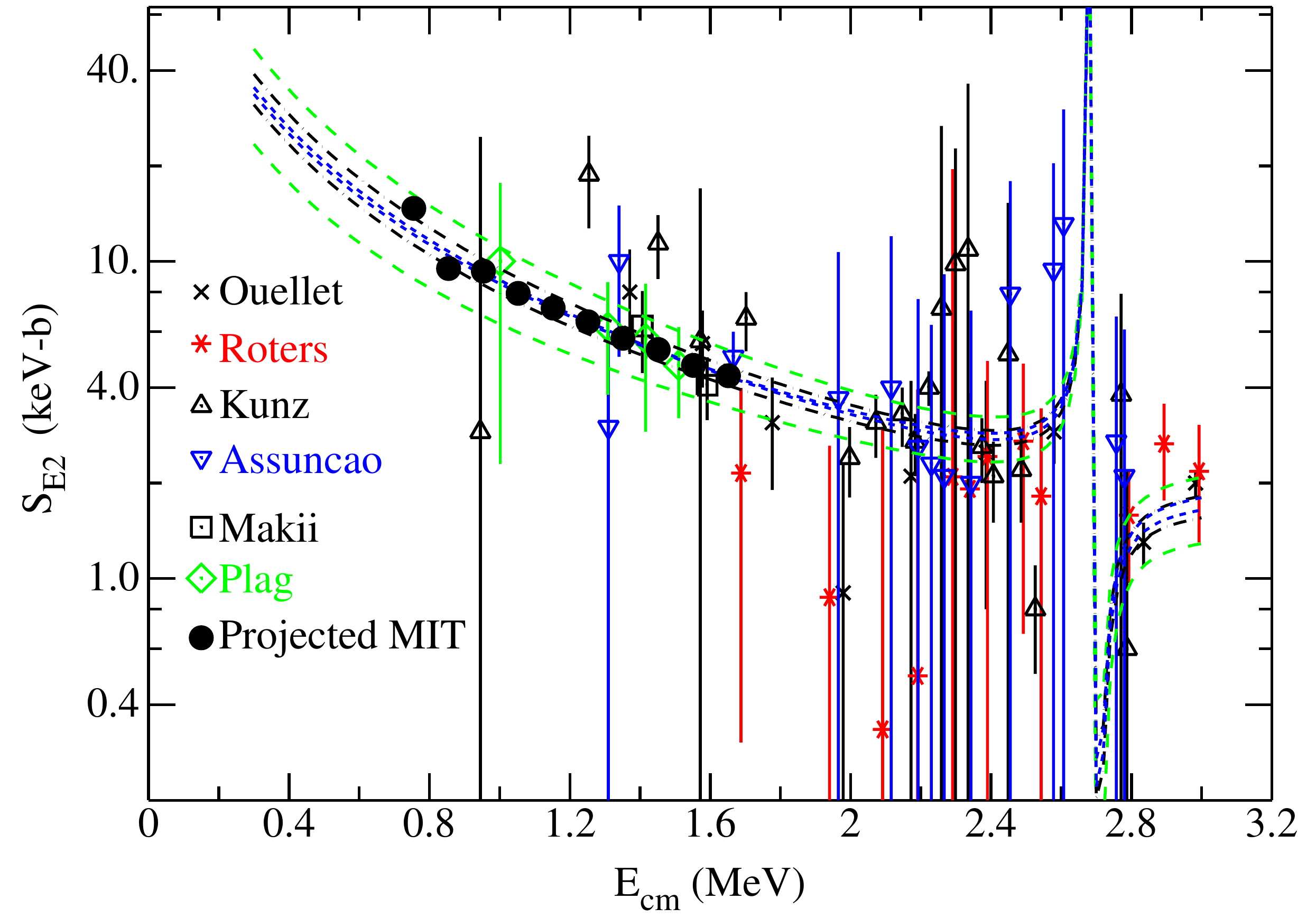}
\caption{The astrophysical S factor for the $E1$ ($E2$) cross section as a function of center of mass energy is shown in the top (bottom) panel.  The dash-dot black lines represent the $\pm \Delta S$ best fit curves and are based on the parameters in Table~\ref{one}, the long dashed outer green lines represent the $\pm 3 \Delta S$ best fit curves, and the short dashed inner blue curves respresent the $\pm \Delta S$ best fit including the projected MIT data\cite{Friscic:2019eow,friscic} shown as the solid black circles and represent the projected data for a 114 MeV incident electron beam, an electron scattering angle of 15$^\circ$ and case A in Ref.~\cite{Friscic:2019eow}.  The existing data are taken from the Refs.~\cite{Dyer:1974pgc,Kremer:1988zz,Redder:1987xba,Ouellet:1992zz,Roters:1999zz,Gialanella:2001ayx,Kunz:2001zz,Assuncao:2006vy,Makii:2009zz,Plag:2012zz}}
\label{fig3}
\end{center}
\end{figure}

 Several observations can be made from Table~\ref{two}.  The standard deviations for the projected S-factors with proposed MIT data are significantly smaller than those without MIT data.   For the fits to the data after 2000, the reduced $\chi^2$ is significantly smaller than that for fits to ``all" data for the $E1$ case, but comparable for $E2$.  This indicates that the $E1$ data sets after 2000 are more consistent with one another than with all data sets.  Finally, the $S$-factor projections for $E2$ are dramatically improved by the projected MIT data.  
As an example, the projections from the fits to all CTAG $E1$ and $E2$ data, the case represented by the first line in Table~\ref{two}, are shown in Fig.~\ref{fig1} and Fig.~\ref{fig2}. 
The dashed vertical line indicates the projection for the fit to the original data, while the histogram represents the results of fits to 1000 sets of randomized pseudo-data.  The dotted curve is a Gaussian based on the mean and standard deviation found from the fits.  

\begin{figure}[h]
\begin{center}
\includegraphics[width=3.8in]{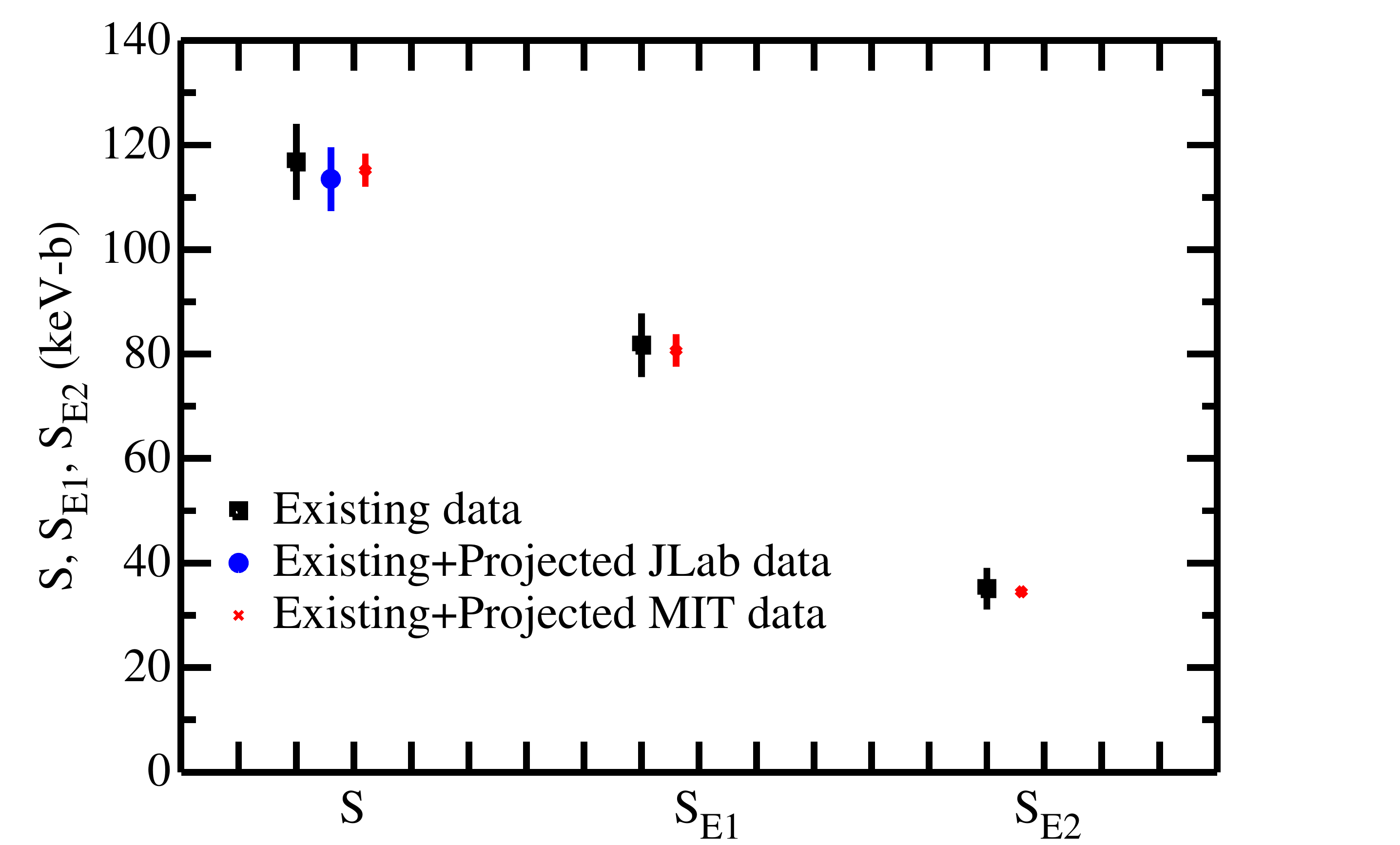}
\caption{Comparison of fit results for existing data (solid squares), including projected JLab data\cite{Holt:2018tbi,suleiman:2014aa} (solid circles), and including projected MIT data (small crosses) for total $S$(300~keV), $S_{E1}$(300~keV) and $S_{E2}$(300~keV).}
\label{fig4}
\end{center}
\end{figure}

Fig.~\ref{fig3} shows the energy dependence of the $S_{E1}$ and $S_{E2}$ factors.  The $S$ factors from the proposed $^{16}$O($e,e'\alpha$)$^{12}$C experiment\cite{Friscic:2019eow,friscic} are shown as the solid green triangles in the figure.  The inner blue short dashes indicate the $\pm\Delta S$ fit that includes the proposed MIT data.  Clearly, the statistical errors are sufficiently small that there is a dramatic reduction expected in the statistical error of the $S$ factor projections to 300~keV.  The $\pm \Delta S$ and $\pm 3\Delta S$ bands are given by the black dash-dot and green long dash curves, respectively, for the case of existing $E1$ and $E2$ data only.   The $E2$ data would be significantly improved with the projected MIT experiment.  

The impact of the new JLab and MIT experiments on the $S$ factors extrapolated to 300~ keV can be readily seen in Fig.~\ref{fig4}.  Furthermore, the projected MIT results extend to lower energy than previous data.
Fig.~\ref{fig4} shows a comparison of fit results for the $S$(300~keV), $S_{E1}$(300~keV) and $S_{E2}$(300~keV) factors for existing data, for existing data including projected JLab data\cite{Holt:2018tbi,suleiman:2014aa}, and existing data including projected MIT data.  
The standard deviation for $S$(300~keV) is somewhat improved by including the projected JLab data.  It is noted that the proposed JLab experiment only measures total $S$.  When the projected MIT data are included in the fits the $E1$ and $E2$ standard deviations as well as the total $\Delta S$ are significantly smaller than that with only existing data.

\section{Summary}
 
From this study it appears that OSEEA reaction data proposed by MIT could have a significant impact on the statistical precision of  S(300~keV).  
The projected standard deviations for the 1000 fits to the $E1$ and $E2$ data with the proposed MIT data is significantly smaller than that without MIT data.   The projected MIT results not only have superior statistical precision, but will also extend to lower energy than previous data.

\section{Acknowledgements}
\begin{acknowledgments}
We thank the late Steven Pieper for significant contributions in the early stages of this work.  We also thank  R. Milner and I. Friščić for useful discussions.  This work is supported by the U.S. National Science Foundation under grant 1812340 and by the U.S. Department of Energy (DOE), Office of Science, Office of Nuclear Physics, under contract No. DE-AC02-06CH11357
\end{acknowledgments}

\bibliography{rmatrix_all_16O_1}

\end{document}